\documentclass[twocolumn,prl,superscriptaddress,bibnotes,notitlepage,nofootinbib]{revtex4-1}
\usepackage{color}
\definecolor{red}{rgb}{0.75,0,0}
\definecolor{blue}{rgb}{0,0,0.75}
\definecolor{green}{rgb}{0,0.5,0}
\usepackage{amsmath}
\usepackage{amssymb}
\usepackage{bm}
\usepackage{graphicx}
\usepackage{verbatim}
\usepackage[T1]{fontenc}
\usepackage{hyperref}
\usepackage{ulem}
\usepackage{lipsum}

\def\be{\begin{equation}}
\def\ee{\end{equation}}
\def\bea{\begin{eqnarray}}
\def\eea{\end{eqnarray}}

\def\besub{\begin{subequations}}
\def\eesub{\end{subequations}}

\def\bwd{\begin{widetext}}
\def\ewd{\end{widetext}}

\newcommand{\bsf}[1]{\textsf{\textbf{#1}}}

\newcommand{\AM}[1]{\textcolor{black}{#1}}
\newcommand{\AMN}[1]{\textcolor{black}{#1}}
\newcommand{\AMNN}[1]{\textcolor{black}{#1}}

\begin{document}
\title{Enhanced orientational ordering induced by active yet isotropic bath}
\author{Ananyo Maitra}
\email{nyomaitra07@gmail.com}
\affiliation{Sorbonne Universit\'{e} and CNRS, Laboratoire Jean Perrin, F-75005, Paris, France}
\author{Raphael Voituriez}
\affiliation{Sorbonne Universit\'{e} and CNRS, Laboratoire Jean Perrin, F-75005, Paris, France}
\affiliation{Sorbonne Universit\'{e} and CNRS, Laboratoire de Physique Th\'{e}orique de la Mati\`{e}re Condens\'{e}e,
F-75005, Paris, France}
\begin{abstract}
Can a bath of isotropic but active particles promote ordering of anisotropic but passive particles? In this paper, we uncover a fluctuation-driven mechanism by which this is possible. Somewhat counter-intuitively,  we show that the passive particles tend to be more ordered upon increasing the noise-strength of the active isotropic bath.
We first demonstrate this in a general dynamical model for a non-conserved order parameter (model A) coupled to an active isotropic field and then concentrate on two examples, i. a collection of polar rods on a substrate in an active isotropic bath and ii. a passive apolar suspension in a momentum conserved, actively forced but isotropic fluid  which is relevant for current research in active systems. 
Our theory,
which is relevant for understanding ordering transitions in out-of-equilibrium systems can be tested in experiments, for instance, by introducing a low concentration of passive rod-like objects in active  isotropic fluids and, since it is applicable to any non-conserved dynamical field, may have applications far beyond active matter.
\end{abstract}
\maketitle
\normalem

Active systems are driven by a direct, isotropic and independent supply of energy at the scale of \emph{individual} constituents, termed active particles which, in dissipating it, perform mechanical work. 
The anisotropy of polar or apolar active particles can non-trivially couple with this \emph{isotropic} energy input leading to counter-intuitive collective properties -- active polar rods on substrates have long-range order in two-dimensions \cite{TT, TT_rean, Sriramrev, TT_Ram} while orientationally ordered phases are rendered unstable in momentum conserved incompressible fluids \cite{Aditi1, RMP}, to mention two examples. Even when active particles do not possess any shape asymmetry, activity leads to spatial clustering and aggregation in the absence of any attractive interaction \cite{MIPS_rev, howfar, MIPS_Cugliandolo}, though, obviously, the system remains isotropic.

While a bath of isotropic active particles can not break rotation symmetry, can they affect the orientational properties of \emph{passive but anisotropic} rods in contact with them? In this paper we show that they can --
%
an active but isotropic bath can induce orientational ordering of passive particles via a fluctuation-driven mechanism for parameter values for which the corresponding passive system would remain isotropic. This activity-driven orientational ordering \emph{increases} with increasing strength of the active noise. We first demonstrate this counter-intuitive effect using the prototypical model for order-disorder transition --
model A \cite{HalpHohen} with a standard scalar $\phi^4$ free energy -- driven by an active autonomously relaxing field, which acts as a nonequilibrium bath. We analytically calculate the shift of the critical point due to activity to first order in the correlation time of the active field and demonstrate that high active noise promotes ordering. We then discuss two examples that are relevant for current research in active systems and may be realised experimentally --\emph{passive polar} rods on a substrate in a bath of active Ornstein-Uhlenbeck particles (AOUPs) \cite{howfar} and apolar rods immersed in a momentum-conserving \AMN{bath of} isotropic swimmers. 
Finally, to justify the upward shift of the critical point, we consider the dynamics of discrete spins in an active bath and demonstrate that the strength of the effective two-particle aligning interaction between spins is enhanced due to the coupling to the active bath.
Our results expose a new active fluctuation-driven mechanism for orientational ordering and indicates that rotation symmetry in active systems may be broken 
\AM{arbitrarily above the \emph{passive} mean-field critical point.}

\emph{$\phi^4$ theory coupled to an active field:} First, let us consider a standard model-A dynamics with a $\phi^4$ free energy coupled to a bath which is modelled as a field that relaxes autonomously:
\begin{equation}
\label{modelAstrMT}
\partial_t\phi({\bf x}, t)=-\frac{\delta F_\phi}{\delta\phi({\bf x})}+c\xi_c({\bf x},t)+\xi_\phi({\bf x},t)
\end{equation}
\begin{equation}
\label{cnoisemodelAMT}
\tau_\phi\partial_t\xi_c({\bf x},t)=-\xi_c({\bf x},t)+\zeta({\bf x},t)
\end{equation}
where $\langle\zeta({\bf x},t)\zeta({\bf x}',t')\rangle=2\tilde{D}_\phi\delta({\bf x}-{\bf x}')\delta(t-t')$, $\langle\xi_\phi({\bf x},t)\xi_\phi({\bf x}',t')\rangle=2D\delta({\bf x}-{\bf x}')\delta(t-t')$, $c$ is an active constant coupling the autonomously relaxing field $\xi_c$ to $\phi$ and
\begin{equation}
\label{fenergphi4MT}
F_\phi=\int d{\bf x} \left[\frac{\alpha}{2}\phi^2+\frac{\beta}{4}\phi^4+\frac{K}{2}(\nabla\phi)^2\right]
\end{equation}
with $\alpha<0$ signalling a mean-field transition to an ordered state and thus, $\alpha=0$ being the mean-field critical point. Typically, the sign of $\alpha$ may depend on a control parameter, which for many active systems, is the particle density \cite{Solon_TT}. However, in this calculation, we simply take $\alpha$ to be a control parameter which can be directly tuned experimentally \cite{fnt1}. 
Activity, which will modify the steady-state distribution from the equilibrium one (when $c=0$) $\propto e^{-F_\phi/D}$, \AM{where $D$ is the noise strength in \eqref{modelAstrMT}}, 
enters \eqref{modelAstrMT} and \eqref{cnoisemodelAMT} through two distinct mechanisms: i. The Onsager symmetry-breaking coupling between $\dot{\phi}$ and $\xi_c$, $\propto c$ and ii. the noise strength \AM{$\tilde{D}_\phi$} in \AM{\eqref{cnoisemodelAMT}} being  unrelated to the damping (i.e., they can be varied independently) and to \AM{$D$} in \AM{\eqref{modelAstrMT}}. Notice that when $\tau_\phi=0$, the coupling to $\xi_c$ in \eqref{modelAstrMT} simply leads to an extra \emph{white} noise and therefore \eqref{modelAstrMT} reduces to the usual model A, albeit one with an enhanced noise strength $\tilde{D}=D+c^2\tilde{D}_\phi$ and the steady-state probability distribution $\propto e^{-F_\phi/\tilde{D}}$. We use this fact to construct the first order in $\tau_\phi$ correction to the steady-state distribution. Crucially, unlike in unified coloured noise approximation (UCNA) \AM{\cite{UCNA}}, we do not ignore the equilibrium white noise in \eqref{modelAstrMT} \cite{fnt2}.

To obtain the steady state distribution for $\phi$, marginalised with respect to $\xi_c$, $\Pi[\phi]=\int \mathcal{D}\xi_c\Pi[\phi,\xi_c]$, we start with the Fokker-Plank equation for $\Pi[\phi,\xi_c]$, and then construct the equation for the moments of this distribution $R_k=\int \mathcal{D}\xi_c\xi_c^k\Pi[\phi,\xi_c]$. We then use the steady-state equations for these moments \AM{to eventually obtain} the steady state distribution $R_0=\Pi[\phi]\propto e^{-\tilde{F_\phi}/\tilde{D}}$ to first order in $\tau_\phi$ (see SI \cite{supp}) where
\begin{widetext}
\begin{equation}
\label{fenrgpert}
\tilde{F}_\phi=\frac{1}{2}\int d{\bf x} \bigg[\{\alpha+\frac{\tau_\phi c^2\tilde{D}_\phi}{\tilde{D}}(\alpha^2-6\tilde{D}\beta)\}\phi^2+\beta\left(\frac{\tilde{D}+4c^2\tau_\phi\alpha \tilde{D}_\phi}{2\tilde{D}}\right)\phi^4+K\left(1+2\frac{\tau_\phi c^2\tilde{D}_\phi}{\tilde{D}}\alpha\right)(\nabla\phi)^2+...\bigg],
\end{equation}
\end{widetext}
with the ellipsis denoting terms higher order in fields and gradients. Importantly, \AM{at $\mathcal{O}(\tau_\phi)$, this model is equivalent to an equilibrium one with a noise-strength $\tilde{D}$ and a free energy \eqref{fenrgpert} (which is perturbative \emph{only} in $\tau_\phi$), and one can, therefore, use the tools developed to treat the classic $\phi^4$ model to discuss its critical propoerties.} Eq. \eqref{fenrgpert} signals an upward shift of the mean-field critical point from $\alpha^0=0$ for $c=0$ to $\alpha^c=6c^2\tilde{D}_\phi\beta\tau_\phi$ to $\mathcal{O}(\tau_\phi)$ due to activity which \emph{increases} with enhanced active noise strength $\tilde{D}_\phi$.  \AMNN{In the passive $\phi^4$ theory, the critical point is depressed below the mean-field one for $2\leq d<4$. However, the active system we consider here can be tuned to criticality by \emph{increasing} the noise strength of the active field coupled to $\phi$.}
\begin{figure}
 \includegraphics[width=7cm]{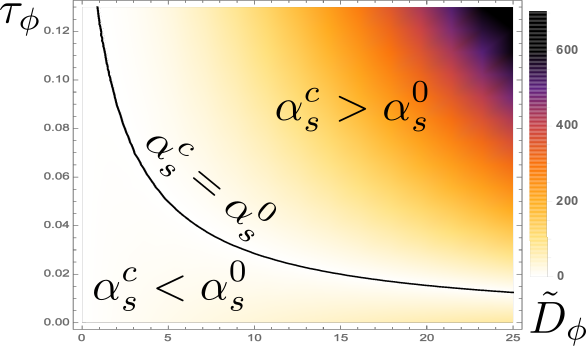}
\caption{The plot of $\alpha^c_s-\alpha^0_s$, the shift of the self-consistent critical point due to the active coupling $c$, as a function of $\tau_\phi$ and $\tilde{D}_\phi$. The intensity of the colours represent $|\alpha^c_s-\alpha^0_s|$. We have taken $\beta=c=K=\mathcal{S}=D=1$ for this plot.} 
\label{Figtranstemp}
\end{figure}
\AMNN{Here, we show this within a self-consistent (Hartree) approximation, explicitly calculating the shift of the critical point in terms of our model's parameters, and argue in the supplement \cite{supp} that this is true more generally.}
If there were no coupling to $\xi_c$ i.e., if $c=0$, the self-consistent critical point would be shifted \emph{downward} to $\alpha^0_s=-3\mathcal{S}D\beta/K$ where $\mathcal{S}=\Omega_d\Delta^{d-2}/[(2\pi)^d(d-2)]$ is a geometric factor with $\Delta$ being an upper wavevector cut-off and $\Omega_d$ being the solid angle subtended by a sphere in $d$ dimensions.
For $c\neq 0$, but in the $\tau_\phi\to 0$ limit, the effect of the active coupling in our model is merely an enhancement of the white noise-strength to $\tilde{D}$ which depresses the self-consistent critical point \emph{even} further: $\alpha^c_s|_{\tau_\phi=0}=-3\mathcal{S}\tilde{D}\beta/K$. However, the first order in $\tau_\phi$ correction to $\alpha^c_s$ is positive implying that the correlation time of the active field can lead to \emph{upward} shift of the one-loop critical point:
\begin{equation}
\alpha^c_s-\alpha^c_s|{_{\tau_\phi=0}}=\tau_\phi3c^2 \tilde{D}_\phi\beta\left(2+\frac{3\mathcal{S}^2\beta\tilde{D}}{K^2}\right).
\end{equation}
This implies that while the mean-field critical point always shifts upward with $\tau_\phi$, the fluctuation-corrected critical point within a self-consistent or random-phase approximation, shifts upward due to the presence of the active coupling $c$ only when 
\begin{equation}
\tau_\phi>\frac{\mathcal{S}K}{2 K^2+3\mathcal{S}^2\beta\tilde{D}}.
\end{equation}
Since $\tilde{D}$ appears in the denominator in this expression increasing $\tilde{D}_\phi$ (or equivalently, $c$) decreases the threshold $\tau_\phi$, beyond which the active coupling leads to an upward shift of the critical point. Thus, at any non-zero value of $\tau_\phi$, one can shift the critical point upward arbitrarily by \emph{increasing} the noise strength of the active field $\xi_c$ (see Fig.\ref{Figtranstemp}). 
This exposes a new active fluctuation-induced mechanism for ordering. However, the coupling to the field $\xi_c$
does not affect the critical \emph{exponents} which are still characteristic of the Wilson-Fisher fixed point; considering a rescaling ${\bf x}\to b{\bf x}$, $t\to b^z t$ and $\phi\to b^\chi \phi$, where $z$ and $\chi$ are the dynamical and roughness exponents respectively, we see $\tau_\phi\to b^{-z}\tau_\phi$. Since the dynamical exponent $z>0$, $\tau_\phi$ must be an irrelevant parameter and must flow to $0$ in any dimension.

\emph{Polar passive particles in an AOUP bath:} We now turn to the description of active systems in which this fluctuation-induced \AM{shift of the mean-field critical point} may be experimentally observed. The first of these consists of polar rods in a bath of AOUPs on a substrate. The polarisation field of the polar rods is described by the two-dimensional vector ${\bf p}$, while that of AOUPs by ${\bf m}$. We assume that the number density of the AOUPs is not conserved (i.e., they can move in and out of the system) to eliminate the effect of long-range interaction that such a conservation law could mediate \cite{Kafri}. Further, we also do not explicitly consider the dynamics of the density of the polar rods since even in the presence of ${\bf m}$, the coupling between density and polarisation cannot change the mean-field critical point \cite{fnt3}. 

The dynamics of ${\bf p}$ to lowest order in gradients is 
\begin{equation}
\label{polp}
D_t{\bf p}=\Lambda{\bf v}-\Gamma_p\frac{\delta F_p}{\delta {\bf p}}+\sqrt{2D\Gamma_p}\boldsymbol{\xi}_p
\end{equation}
where $D_t$ is the co-rotational and advected derivative, ${\bf v}$ is the centre-of-mass velocity of the system composed of polar rods and AOUPs,  $F_p=\int d{\bf x} f_p$,
$
f_p= ({\alpha}/{2}){\bf p}\cdot{\bf p}+({\beta}/{4})({\bf p}\cdot{\bf p})^2,
$
is the free-energy that would control the dynamics in the absence of activity \AM{(since in this case, we only calculate the shift of the mean-field critical point, we only retain the local part of the free energy)} and $\boldsymbol{\xi}_p$ is a unit variance Gaussian white noise. The polarisation field for particles on a substrate also orients along the local centre-of mass velocity ${\bf v}$, and not only its gradient \cite{Ano_pol, Lauga, Harsh, LPDJSTAT} and the strength of this orientational coupling to velocity is given by $\Lambda$. The equation of motion of the overdamped centre-of-mass velocity field is 
\begin{equation}
\label{velp1}
\gamma{\bf v}=\upsilon{\bf m}-\Lambda\frac{\delta F_p}{\delta {\bf p}}+\sqrt{2D\gamma}\boldsymbol{\xi}_v
\end{equation}
where $\boldsymbol{\xi}_v$ is a unit-variance Gaussian white noise and $\gamma$  is the friction coefficient. The term with coefficient $\Lambda$ is an \emph{equilibrium} coupling to the polarisation field required by Onsager symmetry and ${\bf m}$ is the polarisation field of AOUPs which leads to an active force in \eqref{velp1}. 
The polarisation field of AOUPs is assumed not to order and is taken to relax autonomously,
$
\tau_p\partial_t{\bf m}=-{\bf m}+\sqrt{2D_m}\boldsymbol{\xi}_m.
$
\begin{figure}
  \includegraphics[width=7cm]{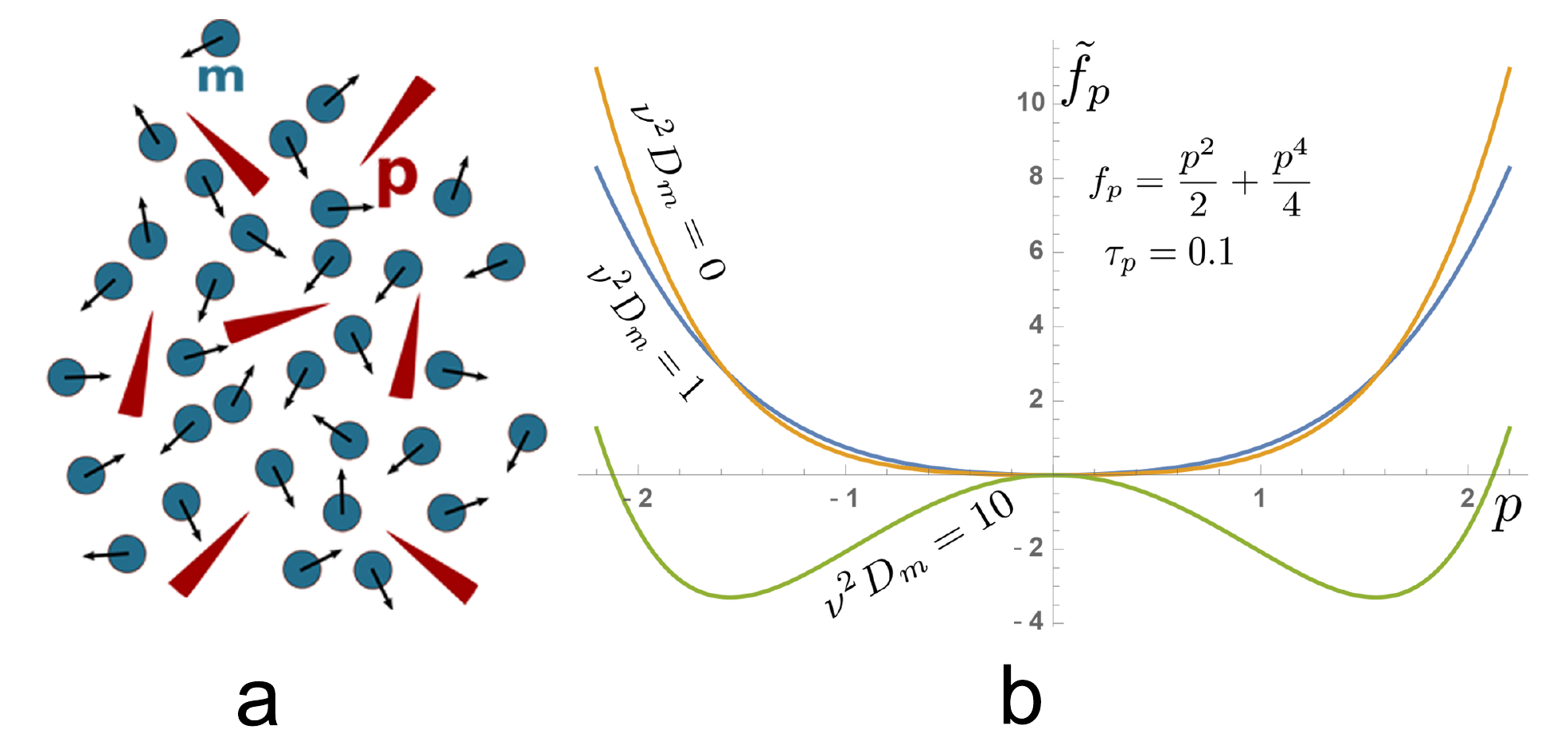}
\caption{(a) Mixture of passive polar rods, denoted by red triangles, and active isotropic particles, denoted by blue circles, with the arrows denoting the instantaneous direction of motion of the active particles, on a substrate. (b) Effective potential $\tilde{f}_p$ for different activity strength $\nu^2 D_m$ ($\tilde{\Gamma}_p=1$ for this figure). This demonstrates that increasing $\nu^2D_m$ leads to a supercritical pitchfork bifurcation with the potential going from having a single minimum at $|{\bf p}|=p=0$  to developing a circle of minima at a finite $p$ and a local maximum at $p=0$.}
\label{Figpot}
\end{figure}
Upon eliminating the velocity field, the coupled equations for ${\bf p}$ and ${\bf m}$ has the form of a vectorial version of the $\phi^4$ model driven by an active bath (with ${\bf m}$ being the bath variable). Therefore, we use the method discussed in that case to obtain the $\mathcal{O}(\tau_p)$ contribution to the steady-state distribution for ${\bf p}$, marginalised over ${\bf m}$, i.e., $\Pi[{\bf p}]=\int d{\bf m}\Pi[{\bf p},{\bf m}]\propto e^{-\tilde{F}_p/\tilde{D}}$ where $\tilde{D}=D+\nu^2 D_m/\tilde{\Gamma}_p$ with $\nu=\Lambda\upsilon/\gamma$ and $\tilde{\Gamma}_p=\Gamma_p+\Lambda^2/\gamma$ and $\tilde{F}_p=\int d{\bf x}\tilde{f}_p$,
\begin{multline}
\tilde{f}_p=f_p+\sum_j\frac{\nu^2 D_m\tau_p}{\tilde{D}}\left[\frac{1}{2}({\partial_p}_jf_p)^2-\tilde{D}{\partial_p}^2_jf_p\right]\\=f_p+\frac{\nu^2 D_m\tau_p}{\tilde{D}}\left[\frac{1}{2}({\partial_p}f_p)^2-\tilde{D}{\partial^2_p}f_p\right]
\end{multline}
where the second equality comes from the fact that $f_p$ only depends on $p=\sqrt{{\bf p}\cdot{\bf p}}$. This demonstrates that increasing the value of $\nu^2D_m$ (which also leads to an increase in the value of $\tilde{D}$) leads to the coefficient of the $p^2$ term in $\tilde{f}_p$ changing sign signalling an \AMN{effectively equilibrium, to $\mathcal{O}(\tau_p)$,} mean-field transition to an ordered state (see Fig. \ref{Figpot}). Further, even if we had introduced the gradient terms in the free energy, and calculated the critical point within a self-consistent theory, the conclusion that enhancing $\nu^2 D_m$ can shift the critical point upwards would remain valid as discussed in the case of $\phi^4$ theory. Thus, this describes an active fluctuation-driven mechanism for ordering of \emph{passive} polar rods in an AOUP bath.

\emph{Apolar rods in a momentum-conserved active isotropic bath:} We now consider \AM{shift of the mean-field critical point} in a system of apolar rods suspended in \AM{a momentum-conserved bath of active but isotropic swimmers in two dimensions} (our calculations remain valid in higher dimensions as well). \AM{As for the polar case, the fluctuations of the density field of the passive apolar rods is ignored since they do not affect the mean-field critical point.} The apolar order parameter that measures the degree of ordering of the passive apolar particles is a traceless, symmetric second-rank tensor ${\bsf Q}$ whose dynamics is
\begin{equation}
\label{apoleq}
D_t{\bsf Q}=\lambda {\bsf A}-\Gamma\frac{\delta F}{\delta {\bsf Q}}+\sqrt{2\Gamma D}\bm{\chi}
\end{equation}
where ${\bsf A}=\nabla{\bf v}+(\nabla{\bf v})^T$ is the strain-rate tensor, with the superscript $T$ denoting transposition, ${\bf v}$ is the velocity of the momentum-conserved fluid, $\lambda$ is the usual flow-alignment parameter \cite{deGen} that implies that shearing a nematic fluid tends to order it along the shearing direction, $\Gamma$ is a dissipative kinetic coefficient, $F=\int d{\bf x}f$ is the free-energy with
$
f= ({\alpha}/{2}){\bsf Q}:{\bsf Q}+({\beta}/{4})({\bsf Q}:{\bsf Q})^2,
$
\AM{(where, as for the polar case, we only retain the local part to calculate the shift of the mean-field critical point)} and $\bm{\chi}$ is a Gaussian white noise of unit variance which would have led to the steady-state probability distribution $\propto e^{-F/D}$ if the model were passive. The equation for the velocity field is
\begin{equation}
\label{veleq}
-\eta\nabla^2{\bf v}=-\nabla\Pi+\lambda\nabla\cdot\frac{\delta F}{\delta {\bsf Q}}+\zeta\nabla\cdot{\bsf M}+
\bm{\Xi}.
\end{equation}
Here, $\eta$ is the viscosity, $\Pi$ is a pressure that acts as a Lagrange multiplier to enforce the incompressibility constraint $\nabla\cdot{\bf v}=0$ and \AM{the noise $\bm{\Xi}$ has the correlator $\langle{\Xi}_i({\bf x},t)\Xi_j({\bf x}',t')\rangle={-2D\eta\nabla^2}\delta({\bf x}-{\bf x}')\delta(t-t')\delta_{ij}$}. Onsager symmetry, which would be operational in the limit of vanishing activity dictates the presence of the force $\propto\lambda$ in the velocity equation. Finally, the velocity field is forced by an active stress ${\bsf M}$ with the coefficient $\zeta$ which models the dipolar forcing due to the bacteria. Since the isotropic part of ${\bsf M}$ cannot affect the velocity field in this incompressible fluid, instead only renormalising the pressure, we can take ${\bsf M}$ to be trace-free without loss of generality. The dynamics of ${\bsf M}$ is not affected by flow since we consider isotropic active particles and is 
$
\tau\dot{{\bsf M}}=-{\bsf M}+\sqrt{2D_M}\bm{\xi},
$
where $\bm{\xi}$ is a Gaussian white noise of unit variance and $\tau$ is the characteristic relaxation time of the active stress.
Fourier transforming and eliminating the velocity field, we find that ${\bsf M}$, which acts as a temporally-correlated active bath enters the (Fourier-transformed) equation for ${\bsf Q}$ the same way as the active field entered the ${\bf p}$ and the $\phi$ equations. Thus, averaging over the directions of the wavevector space (since we consider the isotropic phase and hence there should be no large-scale anisotropy), and following the arguments sketched above and detailed in the supplement, we find an effective steady-state distribution for ${\bsf Q}$, $e^{-\tilde{F}/\tilde{D}}$, marginalised with respect to ${\bsf M}$, where $\tilde{D}=D+\tilde{\zeta}^2D_M/\tilde{\Gamma}$ with $\tilde{\zeta}=\lambda\zeta/8\eta$ and $\tilde{\Gamma}=\Gamma+\lambda^2/8\eta$, and $\tilde{F}=\int d{\bf x}\tilde{f}$,
\begin{equation}
\tilde{f}=f+\tau\frac{\tilde{\zeta}^2D_M}{2\tilde{D}}\left[\frac{1}{2}\partial_{\bsf Q} f:\partial_{\bsf Q} f-\tilde{D}\partial_{\bsf Q}:\partial_{\bsf Q} f\right].
\end{equation}
Using the form of $f$, we again find that increasing $\tilde{\zeta}^2 D_M$ \AM{leads to an instability of} the \emph{disordered} phase. Beyond the limit of stability the disordered phase, \AMN{quasi-long range ordered} nematic state is likely to set in, \AM{which to $\mathcal{O}(\tau)$ is effectively in equilibrium and, unlike orientationally ordered \emph{active} phases in momentum conserved systems \cite{Aditi1, Voit, RMP}, is not generically unstable at large scales} 
-- there is no active stress $\propto {\bsf Q}$ and ${\bsf M}$ does not order along with ${\bsf Q}$, merely contributing a coloured noise to the dynamics of the rotational Goldstone mode. This coloured noise does not \AM{affect the stability of the ordered phase since it cannot modify the sign of the dominant $\mathcal{O}(q^2)$ part of the relaxation rate of angular fluctuations}. 
\emph{Discrete spins coupled to an active field:} To justify the upward shift of the critical point due to activity discussed above we now consider a model of \emph{discrete} orientable particles coupled to an active bath and demonstrate that this coupling leads to an enhancement of the effective two-particle interaction strength.
We assume that the orientable particles interact via the standard spin interaction potential $U=-J\sum_{ij}\cos(\theta_i-\theta_j)$, where the summation may extend over all other spins within a finite radius of a particular spin (the spins are assumed to be static). The dynamics of the spins $\theta_i$ are driven by active autonomous variable $\xi_{c_i}$ as
\begin{equation}
\partial_t\theta_i=-\partial_{\theta_i} U+\chi_i+\xi_{c_i}; \,\,\, \tau_\theta\partial_t\xi_{c_i}=-\xi_{c_i}+\zeta_i
\end{equation}
where $\zeta_i$ are a white noises with variance $2D_a$ and $\chi_i$ are white noises with variance $2D$. The active variables $\xi_{c_i}$ can be thought of as the value of a continuous field $\bm{\xi}_c({\bf x}, t)$ at the position of the $i$-th spin. Thus, this can model a variant of the polar system discussed earlier; fixed spins in a bath of active isotropic particles where the instantaneous polarisation field of the active particles is described by the continuous variable $\bm{\xi}_c$ (the number of the active particles is not conserved, as earlier) and the spins are resolved individually. This leads to a distribution for the spins marginalised with respect to $\xi_{c_i}$: $e^{{-\tilde{U}}/\tilde{D}}$, where $\tilde{D}=D+D_a$ and
\begin{equation}
\tilde{U}=U+\tau_\theta\frac{D_a}{\tilde{D}}\sum_i\left[\frac{1}{2}(\partial_{\theta_i}U)^2-\tilde{D}\partial^2_{\theta_i}U\right]
\end{equation}
The final term, which increases with increasing strength of the active noise, is $-D_a\tau_\theta J\sum_{ij}\cos(\theta_i-\theta_j)$ i.e., it reinforces the strength of the ordering interaction. This reinforcement of the two-body potential ultimately leads to an enhancement of the critical point for the mean-field ordering transition of the spins.
While this considers fixed spins, a similar conclusion should result if the spins themselves are active and move in the direction they point in, as in Vicsek model \cite{Vicsek}. One can heuristically argue for the effective enhancement of the two-particle ordering interaction: consider the limit in which $D_a\gg D$, such that the noise $\chi_i$ can be ignored compared to $\xi_{c_i}$. In this case, the $\xi_{c_i}$ can be eliminated by taking another derivative of the $\partial_t\theta_i$ equation. This leads to an effective friction that is a function of $(\theta_i-\theta_j)$ \cite{howfar}, while the noise remains independent of $\theta_i$ or $\theta_j$. Thus, the ratio of the noise and the friction, which in equilibrium systems is the temperature, becomes a function of $(\theta_i-\theta_j)$. Since this nonequilibrium ``temperature'' is now spin-dependent, the effective distribution becomes a sharper function of $\theta_i-\theta_j$ since the spins also ``cool down'' as they align.

Our calculation \AMN{exposes} a mechanism \AMN{for promoting a transition to an ordered phase in a passive orientable system coupled to an active isotropic one by enhancing the noise of the active bath}. This may provide a possible explanation for ordering in active system that are known to have an oriented state at densities too low for steric interactions to operate \cite{Sano}.
Further, our prediction that polar passive rods in a bath of AOUPs can order above their passive mean-field critical point may be tested experimentally by introducing passive orientable rods in the system of active isotropic particles studied in \cite{Dauchot}. For the purpose of this experiment, passive polar rods will consist of particles with a polar top surface and a circular base instead of a polar base and circular top surface that constitutes active discs. However, care must be taken to isolate our fluctuation-driven effect from the flow-driven one described by \cite{Kafri} (alternatively, the flow-driven effect may be eliminated by randomly adding and removing isotropic active particles). 
Furthermore, the system of apolar particles in an isotropic momentum conserved active fluid that we consider, can model  microtubules in a disordered actomyosin fluid or passive colloidal rods in a solution of spherical bacteria \cite{Rabani}. While actin filaments are themselves orientable and therefore should be affected by flow, in the actively-driven spatio-temporally chaotic state, it is conceivable that their effect may only lead to an isotropic stochastic stress in the force balance equation, albeit one that is both spatially and temporally correlated. In the high activity limit, where the spatial correlations are much smaller than the length-scale of the microtubules, this active noise may be taken to be white in space while being temporally correlated. 
Beyond active systems, our calculation may be applicable in other nonequilibrium contexts such as flow or field driven systems.

\begin{acknowledgments}A.M. acknowledges illuminating discussions with Samriddhi Sankar Ray, Sriram Ramaswamy and Cesare Nardini. This work was supported by ANR grant PHYMAX.
\end{acknowledgments}


\begin{thebibliography}{}
\bibitem{fnt1} Though we do not include the dynamics of the density fluctuations in our discussion, the active shift of the mean-field critical point that we discuss would be valid if one included the dynamics of the density field as well since it only depends on the mean density and not on its fluctuations. Furthermore, for the $\phi^4$ theory, it is valid even at the one-loop order.
\end{thebibliography}
\end{document}